\documentclass[aps,prl,twocolumn,showpacs,preprintnumbers,amsmath,amssymb,groupedaddress]{revtex4}
\usepackage[dvips]{graphicx}
\usepackage{dcolumn}

\begin{document}

\title
{Vortex-lattice melting in two-dimensional superconductors in intermediate fields}

\author{Taro Saiki and Ryusuke Ikeda}

\affiliation{%
Department of Physics, Kyoto University, Kyoto 606-8502, Japan
}

\date{\today}


\begin{abstract} 
To examine the field dependence of the vortex lattice melting transition in two-dimensional (2D) superconductors, Monte Carlo simulations of the 2D Ginzburg-Landau (GL) model are performed by extending the conventional lowest Landau level (LL) approximation to include several {\it higher} LL modes of the superconducting order parameter with LL indices up to six. It is found that a nearly vertical melting line in lower fields, which is familiar within the elastic theory, is reached just by including higher LL modes with LL indices less than five, and that the first order character of the melting transition in higher fields is significantly weakened with decreasing the field. Nevertheless, a genuine crossover to the consecutive continuous melting picture intervened by a hexatic liquid is not found within the use of the GL model. 
\end{abstract}



\maketitle


\section{I. Introduction}

The vortex phase diagram in type II superconductors has been extensively studied in relation to the high $T_c$ cuprates in magnetic fields which is a typical three dimensional (3D) system with strong fluctuation. The vortex lattice melting transition in 3D systems in clean limit has been examined as a first step for understanding phenomena in real superconductors with quenched disorder and is now believed to be of first order in any magnetic field. In contrast, understanding of the phase diagram in 2D case in nonzero magnetic fields have not progressed sufficiently. This is partly because the resulting vortex lattice itself is nonsuperconducting. No state with zero resistance is reached \cite{DSF} in clean limit. Further, a weak but nonvanishing disorder destroys the quasi long range order of vortex positions, and a vortex glass, i.e, a superconducting vortex phase, is never realized at finite temperatures, implying that there will be no phase transition in {\it real} 2D superconductors with disorder in finite fields and finite temperatures \cite{FFH,Nattermann}. 

However, the field-temperature vortex phase diagram in a 2D superconductor remains unresolved even theoretically. 
Based on the elastic theory, there are at least two possibilities, a direct first order melting and the defect-unbinding melting composed of two consecutive continuous transitions and a hexatic liquid crystal phase intervening between them \cite{NH}. In high fields where the pair-field $\psi$ is limited to the modes in the lowest Landau level (LL), however, the 2D melting transition is of first order according to the direct Monte Carlo simulation of the Ginzburg-Landau (GL) model \cite{Kato,JHu}. Then, it will be valuable to clarify whether this first order transition is changed to the consecutive continuous melting scenario in lower fields or not. In fact, the character of the melting transition has been assumed in previous studies to be independent of the strength $H$ of the magnetic field. In addition, when considering the real disordered case, clarifying this issue on the field dependence of the melting mechanism would improve understanding of the 2D vortex states in the following sense: In {\it real} superconducting thin films with {\it weak} quenched disorder, a first order melting in clean limit is not suggested in thermodynamic and resistive data. Note that such experiments are usually performed in much lower fields than $H_{c2}(0)$ where roles of quenched disorder are believed to be weaker \cite{FFH}. The fact that no first order melting has been suggested so far in physical quantities in real 2D thin films might be understood if the melting transition in clean limit is continuous or a highly weak first order one in such lower 
fields. 

Based on such a background on the 2D vortex states, in the present work, the previous numerical analysis limited to the LLL modes of $\psi$ is extended  to the case with several higher LL modes to address the nature of the vortex lattice melting in the GL model in lower fields. 
This paper is organized as follows. In sec.II. the model and the procedures used for the analysis and simulations are explained. In sec.III, the obtained numerical results are explained and discussed. Section IV includes a summary of the present work and a comparison with other theoretical works. 

\section{II. Model and Procedures}

Our starting model, the 2D GL Hamiltonian, takes the form
\begin{eqnarray}
\mathcal{H} &=& s \int d^2 \mathbf{r} [ \varepsilon _0 |\Psi(\mathbf{r})|^2+\xi_0^2|(-i \nabla + 2 |e| \mathbf{A})\Psi(\mathbf{r})|^2 \\ \nonumber
&+& \frac{b}{2}|\Psi(\mathbf{r})|^4 ]
\label{eqn:000}
\end{eqnarray}
with the partition function $Z = {\rm Tr} \exp(-\mathcal{H}/T)$, where $\xi_0$ is the coherence length at $T=0$, $\varepsilon_0 = -1+T/T_{c0}$, $\Psi(\mathbf{r})$ is the pair-field, $s$ is the film thickness, and the Landau gauge $\mathbf{A} = (0, Hx, 0)$ 
will be used hereafter. The magnetic screening due to the {\it fluctuation} of the gauge field ${\bf A}$ will be neglected based on the familiar reasoning that the effective penetration depth \cite{Pearls} $\Lambda=\lambda^2/s$ defined in the Meissner state is, in most cases, beyond the system size, where $\lambda$ is the London penetration 
depth.  

Our simulations have been performed by {\it fixing} \cite{Myojin} the number of vortices $N_s$ in the manner commensurate with the triangular lattice 
\begin{eqnarray}
L_xL_y = 2\pi r_H^2 N_s, \,\,\,\,\,\,\, \frac{L_x}{L_y} =\frac{\sqrt{3}}{2}, 
\end{eqnarray}
where $r_H=(2|e|H)^{-1/2}$, $L_j$ is the system size in each direction, and $N_s^{1/2}$ is assumed to be an integer. Thus, in our simulations, each system size $L_j$ increases with decreasing $H$. Further, the quasi-periodic boundary condition for the pair-field $\Psi(\mathbf{r})$  
\begin{eqnarray}
\Psi(x,y+L_y)&=& \Psi(x,y)\\
\Psi(x+L_x,y)&=& \Psi(x,y) \exp(-i r_H^{-2} L_x y)
\end{eqnarray}
is assumed to be satisfied. Due to this, any gauge-invariant quantities such as $|\Psi(\mathbf{r})|^2$ becomes periodic in the perfect vortex lattice. 

The pair-field $\Psi$ will be expanded in terms of the LL eigenfunctions $\psi_{n,l}$ in the way \cite{Kato,Myojin}
\begin{eqnarray}
\Psi = \sqrt{\frac{T}{s}} \sum _{n=0}^{N_s-1}\sum_{l}c_{n,l}\psi_{n,l}(\mathbf{r}). 
\end{eqnarray}
Then, eq.(\ref{eqn:000}) becomes $\mathcal{{\tilde H}}[bT/s]=\mathcal{H}[bT/s]/T$, where 
\begin{widetext}
\begin{eqnarray}
\mathcal{{\tilde H}}[bT/s] \!\!&=&\!\!  \sum _{n,l} (t-1+(2n+1)h)|c_{n,l}|^2 + \frac{bT}{2 s L_x L_y} \sum _{m_x}\sum _{N_1N_2}\sum_{ \{n_i\} }\sum_{\{l_i\}}
c_{n_1,l_1}c_{n_2,l_2}^*c_{n_3,l_3}c_{n_4,l_4}^* \delta _{l_1+l_3+N_s(N_1+N_2), l_2+l_4}  \nonumber \\  &\times& \mathcal{L}_{n_2,n_1}\left(\frac{k_+}{\sqrt{2}}r_H \right)\mathcal{L}_{n_4,n_3}\left(-\frac{k_+}{\sqrt{2}} r_H \right)  \exp \left[ - \, \frac{k_x^2 + k_y^2}{2}r_H^2 -i \, k_x k_y r_H^2 \right]. 
\label{eqn:002}
\end{eqnarray}
\end{widetext}
Here, $k_+ = k_x + {\rm i}k_y$, and 
\begin{eqnarray}
k_x=\frac{2\pi}{L_x}m_x, \,\,\,\,\,\,\,\, k_y 
= \frac{2\pi}{L_y}(l_1-l_2+N_sN_1)
\end{eqnarray}
with integers $m_x$, $l_1$, $l_2$, and $N_1$. Further, 
the quartic term has been rewritten according to the treatment used elsewhere\cite{HSI} and in terms of the expression 
\begin{equation}
\mathcal{L}_{p, q}(z) = \sum_{n \geq 0}^{{\rm min}(p,q)} \frac{\sqrt{p! q!}}{(p-n)! (q-n)! n!} z^{p-n} (-z^*)^{q-n}.
\end{equation}
In the regime where the phenomenological GL model is applicable, the coefficient $b$ is given, in terms of the GL parameter $\kappa$ and the depairing field $H_{c2}(0)$ at $T=0$, by 
\begin{eqnarray}
b = 8 \pi \frac{\kappa^2}{H_{c2}^2(0)}. 
\end{eqnarray}
As numerical values of material parameters, we use hereafter those typical of optimally-doped high $T_c$ cuprates, $T_{c0} = 10^2 [K]$, $\xi_0 = 10 [\mathrm{\AA}]$, and $s = 15 [\mathrm{\AA}]$. Then, 
$bT/(L_x L_y s)$ is expressed by $W(t,h)/(\pi N_s)$, where 
\begin{eqnarray}
W(h,t) = \kappa^2\cdot10^{-5} \, t \, h, 
\end{eqnarray}
$t=T/T_{c0}$, and $h=H/H_{c2}(0)$. By performing the scale transformation 
\begin{eqnarray}
\{c_{n,l}\} \rightarrow \{c_{n,l}\} /\sqrt{W(h,t)}, 
\end{eqnarray}
eq.(\ref{eqn:002}) finally becomes 
\begin{equation}
\mathcal{{\tilde H}}_{\rm red} = \frac{1}{W(h,t)} \mathcal{{\tilde H}}[(2 r_H^2)^{-1}],
\end{equation}
with $Z = {\rm Tr} \exp(-\mathcal{{\tilde H}}_{\rm red})$ which will be used for numerical simulations. 

As a thermodynamic evidence of a first order transition, we focus on the hysteresis $\Delta E$ of the internal energy, which can be defined by \cite{Myojin}
\begin{eqnarray}
\Delta E= \frac{1}{L_xL_y}\left[ \langle \mathcal{{\tilde H}} \rangle_{\rm dec} - \langle \mathcal{{\tilde H}} \rangle_{\rm inc} \right],
\label{hysteresis}
\end{eqnarray}
where the indices "${\rm dec}$" and "${\rm inc}$" denote cooling and warming processes, respectively, and $\left< \cdots \right>$ implies the thermodynamic average. In the warming process, the vortices is assumed to form the triangular lattice at a low enough temperature. This condition can be expressed as \cite{Kato,Myojin}
\begin{equation}
c_{n,l} = (-1)^{m(m+1)/2} \biggl(\frac{\pi N_s^{1/2}}{\beta_A} \biggr)^{1/2} \delta_{l,0} \delta_{n, N_s^{1/2} m/2}
\end{equation} 
for $0 \leq m < 2 N_s^{1/2}$ but is zero otherwise, where $\beta_A$ denotes the Abrikosov factor ($=1.1596$) of the perfect triangular lattice. On the other hand, the initial condition $c_{n,l} = 0$ 
will be used for the cooling process. 

\section{III. Numerical Results}

In our simulations, the field dependence has been examined at the fixed number of field-induced vortices $N_s$ and by changing $L_j$ following our previous work in LLL \cite{Myojin}, because numerical results become much clearer than those under fixed $L_j$ and a field-induced change of $N_s$. We have performed three simulations with $N_s=36$ and six LLs ($0 \leq n \leq 5$), $N_s=64$ and the six LLs at only $h=0.1$ and $0.3$, and $N_s=36$ and seven LLs ($0 \leq n \leq 6$). We have not found any $N_s$-dependent essential differences in the hysteresis data and snapshots of the vortex configurations. Hereafter, we will primarily show the data resulting from the use of six LLs and $N_s=36$. 

Simulations have been performed by creating Markov chains of the coefficients $c_{n,l}$ according to the Metropolis algorithm. MonteCarlo (MC) steps of the range between $5.0 \times 10^5$ and $1.0 \times 10^6$ were used to ensure approach to the thermodynamic equilibrium, and, afterward, additional $5.0 \times 10^4$ MC steps were used to take a statistical average of physical quantities. 

\begin{figure}[b]
\scalebox{0.3}[0.3]{\includegraphics{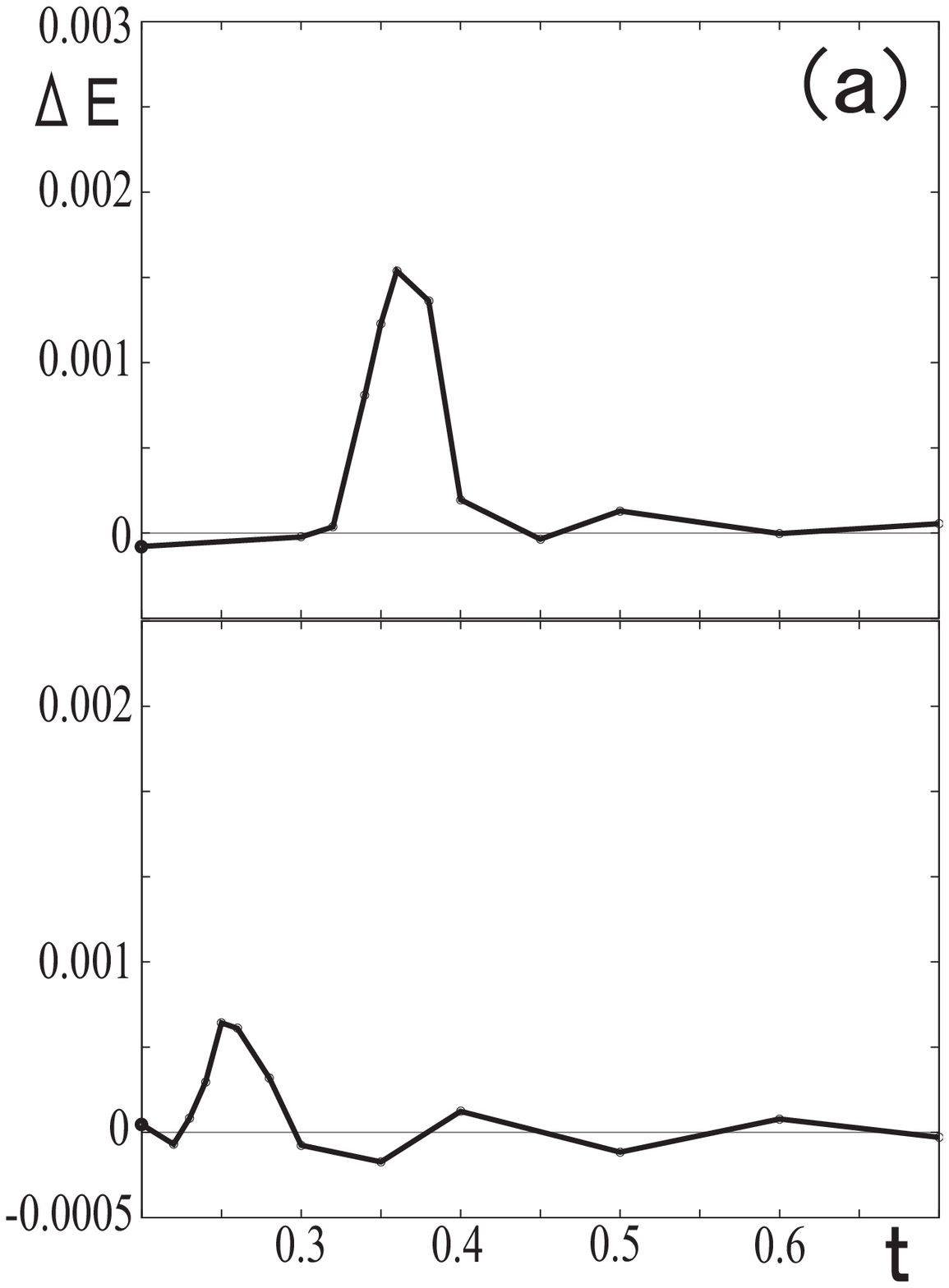}}
\scalebox{0.3}[0.3]{\includegraphics{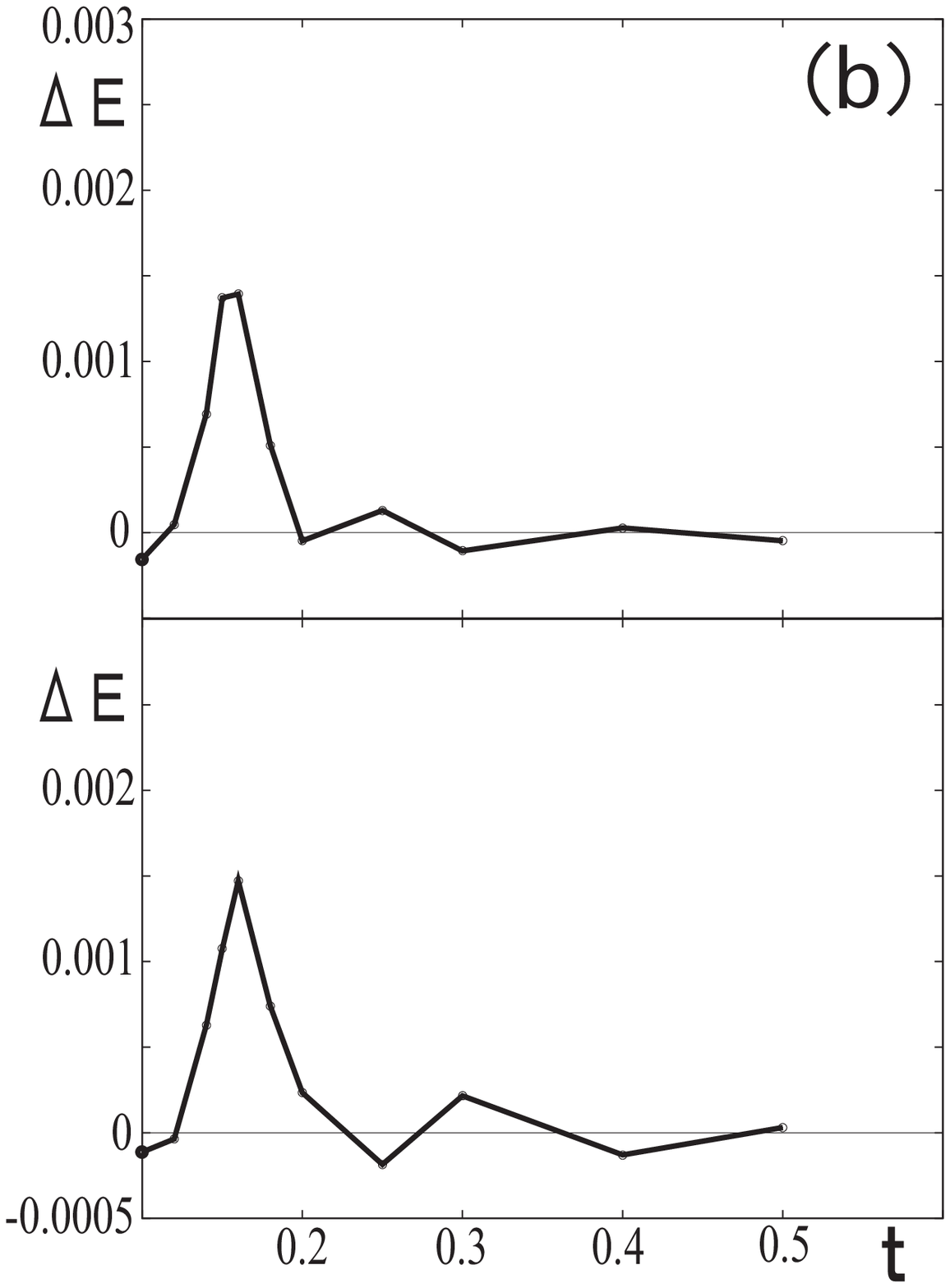}}
\caption{Numerical $t$ ($=T/T_{c0}$) v.s. $\Delta E$ curves in (a) $h=0.3$ and (b) $h=0.6$ taken at the fixed $N_s=36$. For both, the upper figure is the result in the LLL approximation, while the lower one is obtained by taking account of the six LLs with $0 \leq n \leq 5$.}
\end{figure}
Now, let us explain our numerical results. Figure 1 expresses the hysteresis, eq.(\ref{hysteresis}), accompanying the first order melting transition at $t_m(h) \equiv T_m(H)/T_{c0}$ at two different magnetic fields, (a) $h=0.3$ and (b) $h=0.6$, on sweeping the temperature in the LLL approximation (upper figures) and in the case with higher LLs with $n \leq 5$ (lower ones). 
It is found that, compared with the familiar LLL results, inclusion of higher LLs depresses $t_m$ and reduces the hysteresis around $t_m$. In $h=0.3$, these higher LL effects are clearly seen, while the LLL approximation seems to be valid even quantitatively in $h=0.6$. 

\begin{figure}[t]
\scalebox{0.5}[0.5]{\includegraphics{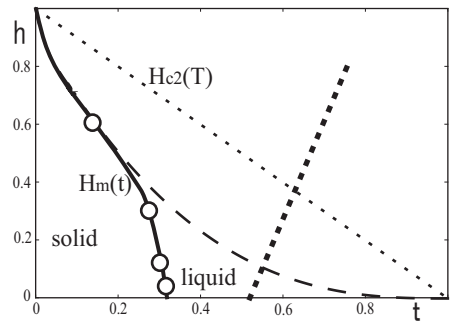}}
\caption{Resulting melting transition curve $t_m(h)$ (solid curve with open circles) from the simulation with six LLs incorporated. The dashed curve is the corresponding curve in the LLL approximation. The thin dotted curve is the $H_{c2}(T)$ line, while the thick dotted curve is the estimated boundary above which the thermally-induced vortex-pairs appear.}
\end{figure}

Prior to a further discussion on the field dependence of the hysteresis, higher LL effects on the $h$-$t$ phase diagram will be explained here. In Fig.2, the melting transition curve (solid curve with open circles) following from the present work is shown together with the corresponding result (dashed curve) in the LLL approximation. The melting temperature at each $h$-value has been defined as the $t$-value at which the hysteresis becomes maximal. In obtaining these curves, we have used the value $\kappa=61$ for the GL parameter. 
In LLL approximation, the reduced melting temperature $t_m$ yields the LLL scaling $1-t_m-h \propto (t_m \, h)^{1/2}$ \cite{RI95}, and the results in Fig.2 show that  
\begin{equation}
1 -t_m -h = c_2^{-1} \sqrt{\frac{W(t_m,h)}{\pi}} 
\label{LLLsim}
\end{equation}
with \cite{Kato} $c_2=0.0989$. The r.h.s. of eq.(\ref{LLLsim}) is proportional to $(h t)^{1/2}$, and the resulting $t$-$h$ relation is called the LLL scaling \cite{RI95}. 
On the other hand, in low enough fields, the 
relation 
\begin{equation}
1 - t_m - h = \frac{W(t_m, h)}{2 \pi c' h} 
\label{HLLsim}
\end{equation} 
is expected to be satisfied \cite{RI95}. Reflecting the fact that the r.h.s. of eq.(\ref{HLLsim}) is $h$-independent, the resulting melting curve in lower fields is nearly vertical in the $t$ v.s. $h$ phase diagram. One can verify that, in Fig.2, the weak field dependence of the solid curve in such low fields is a reflection of the field dependence of $T_{c2}(h)$ rather than the neglect of other higher LLs ($n \geq 6$). In contrast to the LLL approach, however, there is no well-established value of $c'$ in eq.(\ref{HLLsim}). 
Figure 2 suggests that $t_m = 0.3$ in low enough fields, which implies that 
$c'=0.0025$. Namely, the relation $c'=0.25 c_2^2$ is approximately satisfied in the results shown in Fig.2. 

Here, we will compare the solid curve in Fig.2 with the corresponding one following from the elastic theory. To do this, the results in the elastic theory \cite{DSF,Doniach} should be reviewed. In any approach based on the elastic energy of the vortex lattice, the field dependence of the melting temperature $T_m(H)$ is determined by that of the shear elastic molulus $C_{66}$, which depends on the magnitude of the {\it reduced} applied field $h \equiv H/H_{c2}(0)$, where $H_{c2}(T)$ is the depairing field in the orbital limit. even in the low field regime where the phase-only model is useful. Then, $C_{66} \propto H$ in lower fields, while $C_{66} \sim (H-H_{c2})^2$ in higher fields \cite{Brandt}. Consequently, $T_m(H)$ is $H$-independent in low fields, while $T_m(H)$ obeys the lowest LL scaling \cite{DSF}. 
Clearly, the low field portion of the solid curve in Fig.2 is comparable with the $H$-independent $T_m(H)$-curve in the London limit. 
The above-mentioned $t_m(H)$-curve in the elastic theory can be described by a {\it single} expression 
\begin{equation}
\frac{C_{66} r_H^2}{T_{c0}} = \alpha t_m
\label{Lindemann}
\end{equation} 
implying a comparison between the elastic and thermal energies 
by assuming that the mechanism of the vortex lattice melting is universal and uniquely given irrespective of the magnetic field strength, 
where the coefficient $\alpha$ is $h$-independent. The vortex shear modulus $C_{66}$ is expressed, in the case of the triangular lattice, as $H_{c2}(T) H/(32 \pi \kappa^2)$ in the low field London regime and $0.708 (H_{c2}(T) - H)^2/(32 \pi \kappa^2)$ in the high field LLL regime \cite{Brandt}, where the value $\beta_A=1.1596$ for the triangular lattice was used. Then, we find that the elastic model leads to the relation $c'=0.704 c_2^2$ when $t_m(h)$ in the above-mentioned two regimes is expressed by eqs.(\ref{LLLsim}) and (\ref{HLLsim}). That is, the present $t_m(H)$-result shown in Fig.2 suggests that the melting temperature in low fields where the LLL approximation breaks down is over-estimated in the elastic theory. The main origin of this discrepancy consists in the presence of higher LL {\it fluctuations}, which are not included in the elastic theory, in the present numerical simulations of the GL model. Since the mean field solution of the triangular or hexagonal vortex lattice is described only by the LLs with LL indices of multiples of six, reflecting its six-fold orientational symmetry \cite{Lasher}, the shear modulus in this state is also determined by those LLs \cite{RI92}. In other words, the higher LLs with $1 \leq n \leq 5$ are not incorporated in the relation (\ref{Lindemann}). On the other hand, the $n=6$-LL is not included in our computation leading to Fig.2, while the mean-field GL theory should reduce to the London theory just by incorporating the higher LL modes with LL indices of multiples of six. 
Namely, the nearly vertical $t_m(h)$-line in lower fields in Fig.2 has been obtained irrespective of reduction to the London model due to the lowering of $h$. In other words, the elastic model overlooks crucial fluctuation effects and is not sufficient for describing the vortex lattice melting as far as choosing the GL model as the starting model is valid. 

\begin{figure}[b]
\scalebox{0.3}[0.3]{\includegraphics{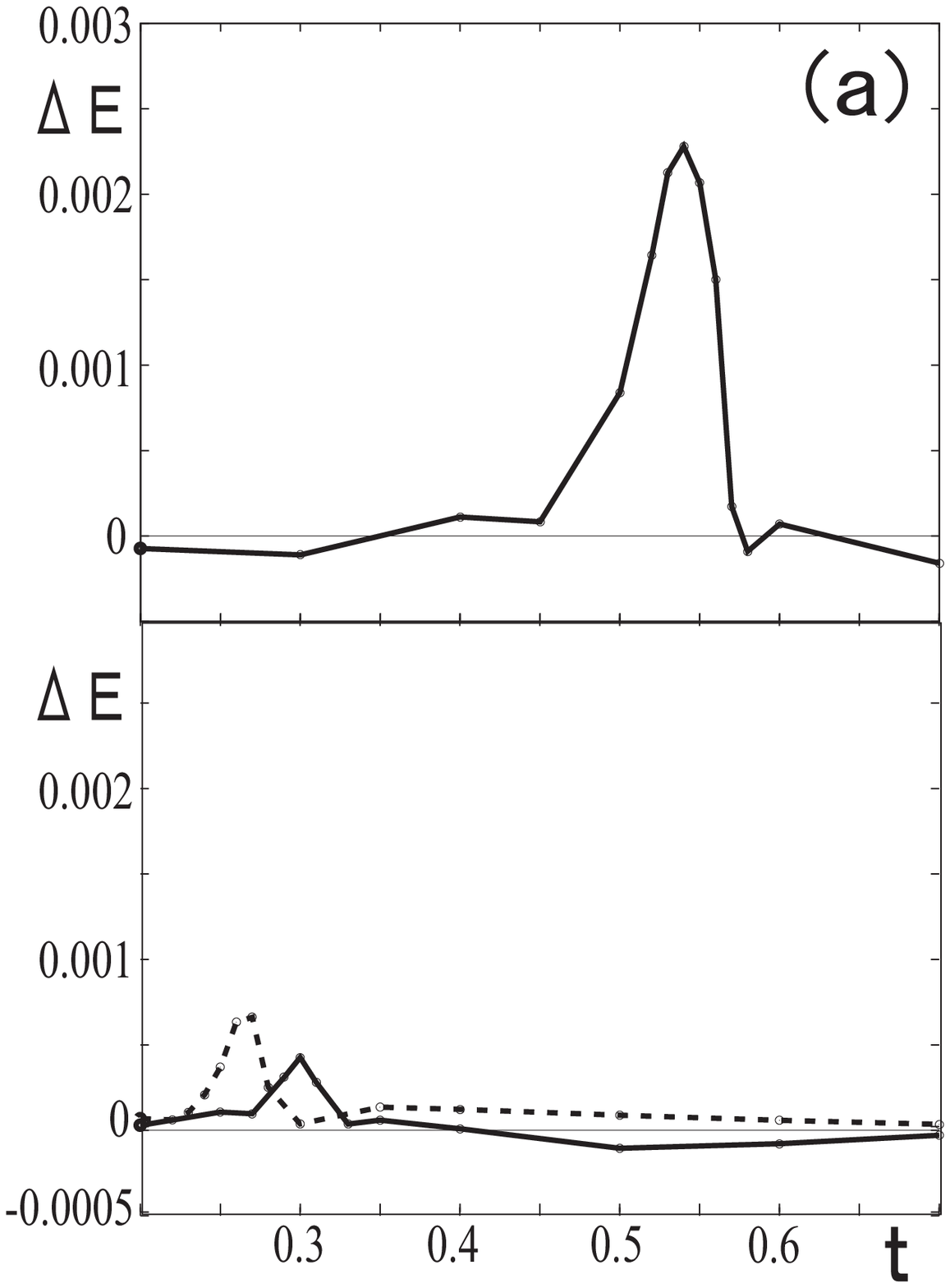}}
\scalebox{0.3}[0.3]{\includegraphics{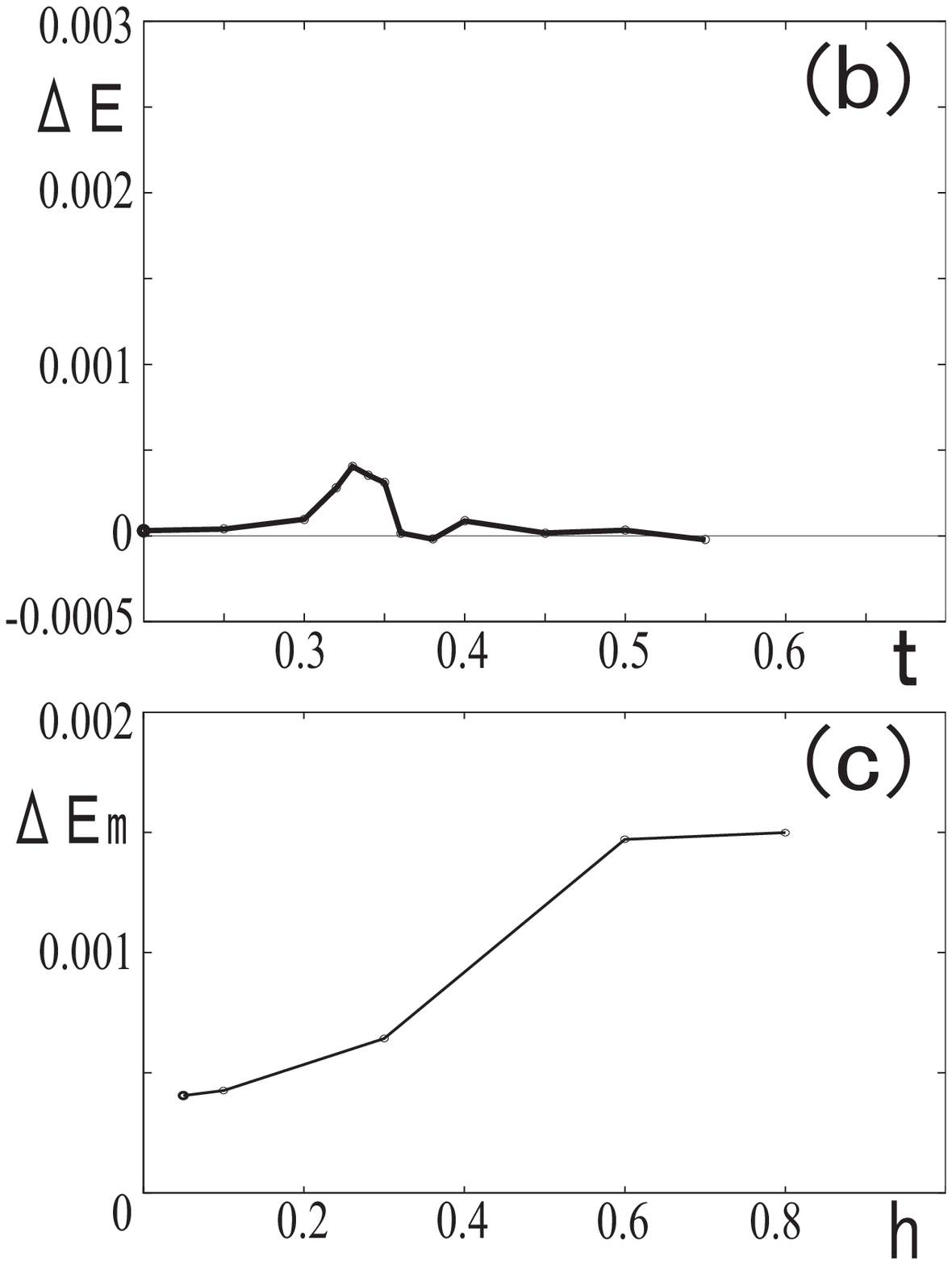}}
\caption{Numerical $t$ v.s. $\Delta E$ data at (a) $h=0.1$ and (b) $h=0.05$. The upper figure of (a) is the result in the LLL approximation, while the lower one of (a) includes the solid curve in the case with six LLs ($n \leq 5$) and the dashed one in the case with seven LLs ($n \leq 6$). The figure (c) implies the $h$-dependence of $\Delta E_m \equiv \Delta E(t=t_m)$. Data in (b) and (c) result from the use of six LLs with $n \leq 5$. }
\end{figure}
Returning to Fig.1, one can see that the magnitude of the hysteresis in the simulations including higher LLs has a different field dependence from that in the LLL approximation. The hysteresis curves following only from LLL indicate that, as is seen from the upper figures in Figs.1 and 3, the hysteresis accompanying the transition increases with decreasing $h$. Once the higher LLs are incorporated, however, the hysteresis peak at the transition rather decreases with decreasing $h$, implying a reduction of the first order character of the melting transition in lower fields. Of course, this field dependence correlates with the corresponding $h$-dependence of the difference in the $t_m$-value between the LLL curve (dashed curve) and the solid one. However, this trend that the first order transition is weakened by decreasing the field is interesting in that it suggests the possibility that, in lower fields, the first order melting might be transmuted to the two-step continuous melting scenairio \cite{NH}. To clarify this possibility within our GL study, the corresponding hysteresis curves in lower fields are shown in Fig.3 together with the $h$-dependence of the peak height of hysteresis at $t_m$ (Fig.3(a)). Although the hysteresis reduces with decreasing field even for lower $h$ ($=0.1$ and $0.05$), the data suggest that the reduction of hysteresis saturates with vanishing $h$. This saturation of the peak height does not seem to be an artifact due to the limitation to the LLs with $n \leq 5$: An additional (dashed) hysteresis curve for $h=0.1$ obtained by including the $n = 6$ LL further is presented in Fig.3 (a). Although the transition point has been slightly shifted to a lower temperasture by the $n=6$ LL which, as mentioned earlier, affects the elastic energy of the mean field vortex lattice, the hysteresis at $t_m$ is slightly bigger by including the $n=6$ LL. Therefore, based on the present simulation, we argue that the melting transition remains of first order even in low field limit, although its first order character is significantly weakened with decreasing the field. 

\begin{figure}[t]
\scalebox{0.4}[0.4]{\includegraphics{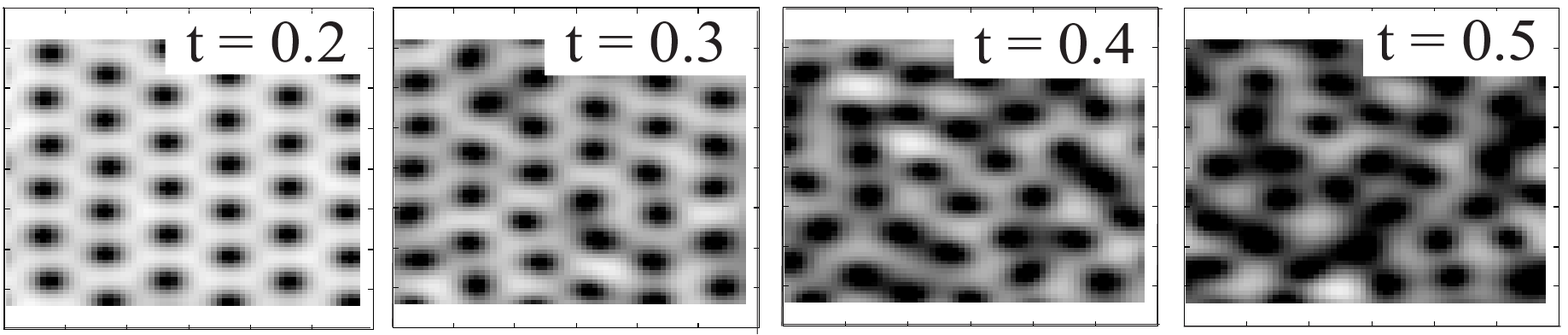}}
\scalebox{0.35}[0.35]{\includegraphics{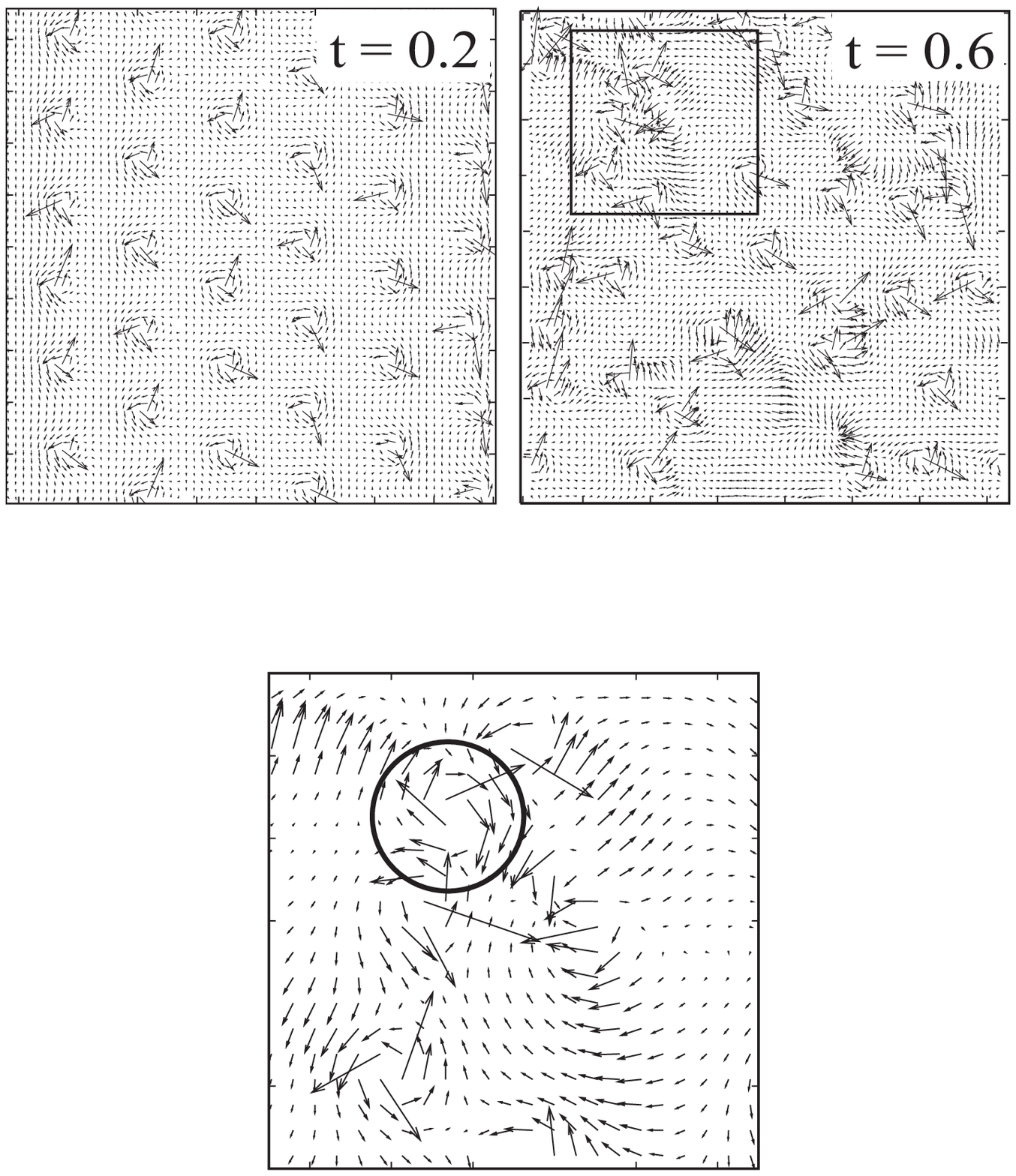}}
\caption{Snapshots of the spatial distribution of $|\Psi({\bf r})|$ (top figures) and the gauge-invariant gradient vector, eq.(\ref{gauge}), (middle and bottom ones) at each $t$ indicated in the figures. The bottom figure is the zoom of the region specified by a square in the middle one at $t=0.6$ and shows the presence of one antivortex (circle) with a clock-wise circular current in contrast to others with a counterclock-wise circular current. }
\end{figure}
Next, we explain the vortex states at various $t$-values seen in our simulation using the LLs $0 \leq n \leq 5$. In Fig.4, we have shown snapshots of the distributions of the order parameter amplitude $|\Psi|$ (upper figures) and the gauge-invariant gradient 
\begin{equation}
\nabla \Phi + \frac{2 \pi}{\phi_0} {\bf A}
\label{gauge}
\end{equation}
of the phase $\Phi$ of $\Psi$ (middle and lower ones). Before discussing what the figures imply, the relation between $\Psi$ and the vortices in the LLL approximation will be reviewed. As is well known, the order parameter within the LLL satisfies \cite{Tesanovic}
\begin{equation}
\Psi_{\rm L}({\bf r}) = \exp\biggl(- \frac{x^2}{2 r_H^2} \biggr) 
f_{N_s}(y+{\rm i}x)
\label{LLL}
\end{equation}
in the present gauge ${\bf A}=H x {\hat y}$, where the function $f_m(z)$ is any $m$-th order polynomial of $z$. Thus, $\Psi_{\rm L}$ has $N_s$ zero points ${\bf r}_\nu=(x_\nu, y_\nu)$ and takes the product form $\Pi_\nu [y-y_\nu + {\rm i}(x - x_\nu)]$ 
Then, it is staightforward to show that the phase $\Phi$ of $\Psi_{\rm L}$ satisfies the topological condition 
\begin{equation}
\nabla \times \nabla \Phi = 2 \pi \sum_{\nu=1}^{N_s} \delta(x-x_\nu) \delta(y-y_\nu) {\hat z},
\label{topo}
\end{equation}
implying each zero point of $|\Psi_{\rm L}|$ is a vortex coordinate. Equation (\ref{topo}) implies that, in $\Psi_{\rm L}$, all of the $N_s$ zero points of $|\Psi_{\rm L}|$ corresponds to the field-induced vortices and thus that no antivortices can appear. That is, the thermally-induced vortex pairs are described not by the LLL modes but only by the higher LL modes of $\Psi$ \cite{RI95}. Bearing this role of the higher LLs in mind, one finds in Fig.4 where $h=0.1$ that, below $t=0.5$, the antivortices do not appear. As is seen in the 
bottom figure, appearance of an antivortex and thus, of one vortex pair induced by the thermal fluctuation is verified just above $t=0.5$. By similarly defining the temperature, at which the thermal vortex-pairs begin to appear, at different $h$-values, we obtain the crossover line, shown in Fig.2 as the thick dotted line, which separates the vortex liquid composed only of the field-induced vortices from the state with the thermally-induced antivortices. This result is consistent with the picture argued elsewhere \cite{RI95} that the higher LLs with antivortices included become more important at higher temperatures and push the melting transition curve down to lower temperatures (see Fig.2 (a) in Ref.\cite{RI95}). 

Finally, we note that, as is seen at the top of Fig.4, the snapshot taken just at $t_m(h=0.1)=0.3$ does not include any dislocation. This is strange at least within the conventional picture based on the elastic theory that the melting transition is driven, more or less, by thermally-induced dislocations \cite{NH}. From the figure, we feel that spatial variations of the amplitude $|\Psi|$ of the pair-field assist the nearly harmonic shear elastic modes and that their cooperative roles result in the melting of the vortex lattice. This view arguing the necessity of the amplitude fluctuation does not contradict the conventional wisdom that Goldstone modes in an ordered phase cannot become critical modes for driving a thermal disordering. Nevertheless, we have to mention that it is beyond the scope of the present work to judge whether such a role of the fluctuation of the amplitude $|\Psi|$ is essential or an artifact of the use of the GL model which is usually valid when $|\Psi|$ is small.

\section{IV. Summary and Discussion}

In the present work, numerical simulations of the GL model with not only the order parameter modes in LLL but also those in five or six higher LLs have been performed. It has been found that the nearly vertical melting curve, which has been identified so far with the result following from the elastic model in the London limit, is obtained simply by incorporating the lower five higher LLs which do {\it not} contribute to the elastic model, and that the first order character of the melting transition diminishes with decreasing the field, although the melting picture with two continuous transitions \cite{NH} is not reached even in the low field limit. The weaker first order transition in lower fields suggests that the discontinuous nature of the transition in clean limit is easily lost by a weak pinning effect in real systems with quenched disorder and thus, explains why the first order melting transition has not been reflected, e.g., in transport data in real superconducting films in nonzero field \cite{RI96}. 

In contrast to the present result in the GL model showing a small but nonzero hysteresis at the melting transition in any magnetic field, 
a simulation work \cite{Burkov} has recently been reported indicating a continuous melting transition. There, the authors have discretized the relation $u_j = r_H^2 ({\hat z} \times \nabla \delta \Phi)_j$, justified \cite{Moore,RI92} within the {\it linear} elastic theory, between the shear displacement ${\bf u}$ and the phase fluctuation $\delta \Phi$ to invoke a starting lattice model for their numerical studies. However, their starting model includes phase slips in the $j$-direction and the direction perpendicular to this on the same footing, suggesting that thermally-induced vortex pairs coexist with the thermally-induced dislocation pairs. That is, the model in Ref.\cite{Burkov} is not compatible with the GL model in which the vortex pairs never appear close to the melting transition (see Fig.2), and thus, it is not surprising that the present results are not consistent with the continuous transition \cite{Burkov} found in a model which, in our opinion, is incompatible with the original GL model. 

On the other hand, we do not definitely conclude at this stage the absence of the low field regime in which the melting transition is continuous \cite{NH} with no thermally-induced vortex pairs. More or less, the main drawback of the present work is the use of the GL model in addressing the low field regime. In such low fields and low temperatures, the amplitude $|\Psi|$ is rigid enough to justify the phase-only model (with no thermally-induced vortex pairs). On the other hand, the fact that the nearly vertical melting curve in low fields expected from theoretical arguments is obtained simply by taking account of five or six higher LLs' modes suggests that the results are not significantly affected by the truncation of the number of incorporated LLs. Nevertheless, further theoretical progress will be necessary to resolve this issue 
on the character of the transition.

This work was supported by Grant-in-Aid for Scientific Research [No. 21540360] from MEXT, Japan.

\end{document}